# Multidimensional Soliton Systems: an Update

Boris A. Malomed[1,2]

[1]Department of Physical Electronics, School of Electrical Engineering, Faculty of Engineering, Tel Aviv University, Tel Aviv 69978, Israel

[2]Instituto de Alta Investigación, Universidad de Tarapacá, Casilla 7D, Arica, Chile

**Abstract**

A briefly formulated update of the recently published review [1] on the topic of multidimensional solitons (chiefly, in nonlinear optics and atomic Bose-Einstein condensates (BECs)) is presented. The update briefly summarizes some notable results on this topic that have been reported very recently, and offers a compact outline of the directions of the current experiment and theoretical work in this and related fields. In particular, as concerns newest experimental findings, included are the creation of multiple *quantum droplets* in prolate BEC, and the observation of expanding toroidal light structures in linear optics. The update of theoretical results includes the analysis of 2D solitons in the optical system with the quadratic nonlinearity and *fractional diffraction*, and the prediction of stable 3D *vortex solitons* in various BEC schemes with long-range interactions.

This brief article offers an update to the recent paper [1], which provided a concise review of the vast topic dealing with the creation, stability, dynamics, and potential applications of two- and three-dimensional (2D and 3D) solitons and soliton-like states, such as self-trapped modes with embedded vorticity, and *quantum droplets* (QDs) in Bose-Einstein condensates, stabilized against the mean-field (MF) collapse by beyond-MF effects of quantum fluctuations [2]. Review [1] addressed both theoretical and experimental aspects of the topic. It was an up-to-date condensed compendium of a broad presentation of the theme of multidimensional solitons reported three years ago in book [3]. This theme finds diverse realizations in two general fields, *viz*., optics and studies of matter waves in atomic Bose-Einstein condensates (BECs).

Although review [1] was published less than two years ago, the rapid development has produced new findings in the experiment and theory, which makes it appropriate to present the current update. As concerns the experimental creation of QDs in BECs, it is relevant to mention an essential addition to the previous results, which reported the creation of multiple quasi-2D [4] and 3D droplets [5] via development of the capillary instability of elongated binary BEC clouds, leading to the fission into two or several stable QDs [6]. The original elongated cloud was prepared in the non-interacting BEC, loading it into a cigar-shaped (quasi-1D) trapping potential; then, the self-focusing nonlinearity was suddenly switched on (i.e., the *quench* was applied) by means of the Feshbach resonance, which makes atomic collisions effectively attractive under the action of properly chosen spatially uniform dc magnetic field). The simplest example of the quench-initiated splitting of the primary elongated cloud into a couple of stable QDs is displayed in Fig. 1.

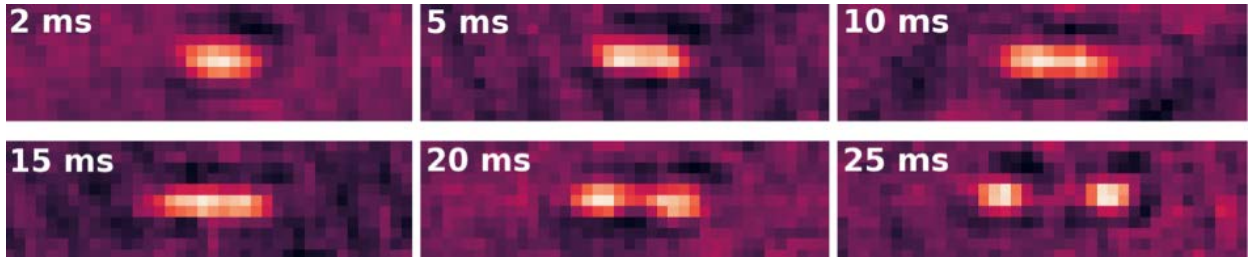

Fig. 1. The spontaneous splitting of the primary cloud of ultracold $^{41}$K atoms into a pair of stable QDs [6], displayed as a sequence of atomic-density patterns observed at different values of, as indicated in the panels.

In optics, recent experiments have also revealed spatially-confined states, which are similar to 3D solitons, including ones with a complex inner topological structure, such as *skyrmions*, see a recent review [7] and an example in the form of a toroidal optical pulse reported in [8], which is shown in Fig. 2. However, these modes, although being quite interesting, are not solitons, as they are produced by the linear light propagation, hence they gradually expand, without exhibiting any self-trapping. A large number of new results for the spatiotemporal optical structures in linear and nonlinear settings are collected in the recent special issue of journal Nanophotonics [9].

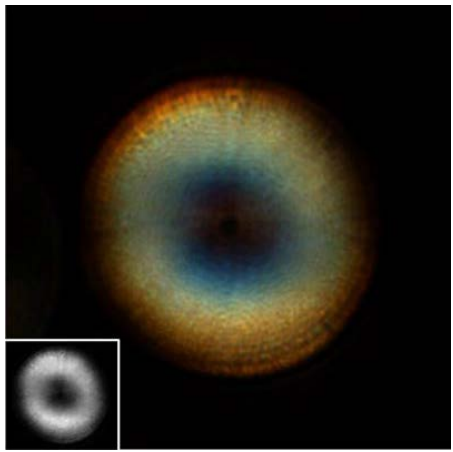

Fig. 2. A snapshot of an expanding toroidal optical mode observed in the course of the linear light propagation [8]. The color in the main plot designates a local dominant value of the wavelength of light, while the inset displays the power-density profile. Unlike 3D vortex solitons in nonlinear media, which demonstrate a similar distribution of the local power [1,3], this linear mode carries no optical angular momentum.

Many new theoretical results have been published very recently, developing the known concepts and putting forth new ones in the studies of multidimensional solitons in optics and BECs. Also, some newly published reviews summarize theoretical (and partly experimental) results in this research area, which were not properly summarized previously. In particular, these are reviews on the topic of vorticity-carrying QDs in BEC [10] and multidimensional solitons in discrete and semi-discrete media (the semi-discreteness implies that the system is discrete in one or two directions, being continuous in the remaining one(s)) [11].

A new direction is the study of multidimensional solitons in optical media with an effective *fractional diffraction*. A relevant example is the 2D fractional system with the quadratic (second-harmonic-generating, alias $\chi^{(2)}$) nonlinearity [12], which is based on the following system of propagation equations for amplitudes *U* and *W* of the fundamental and second-harmonic waves:

$$i\frac{\partial U}{\partial z} = \frac{1}{2}\left(-\frac{\partial^2 U}{\partial x^2} - \frac{\partial^2 U}{\partial y^2}\right)^{\frac{\alpha}{2}} U - WU^*, \qquad (1)$$

$$2i\frac{\partial W}{\partial z} = \frac{1}{2}\left(-\frac{\partial^2 U}{\partial x^2} - \frac{\partial^2 U}{\partial y^2}\right)^{\frac{\alpha}{2}} W + QW - \frac{1}{2}U^2, \qquad (2)$$

where *z* is the propagation distance, (*x,y*) are the transverse coordinates, positive $\alpha < 2$ is the *Lévy index*, which determines the fractality of the system, the asterisk stands for the complex conjugate, and real *Q* is the mismatch parameter. The fractional diffraction operator in Eqs. (1) and (2) is actually an integral one, defined as the 2D *Riesz derivative*,

$$\left(-\frac{\partial^2 U}{\partial x^2} - \frac{\partial^2 U}{\partial y^2}\right)^{\frac{\alpha}{2}} U = \frac{1}{(2\pi)^2}\int_{-\infty}^{+\infty}dp\int_{-\infty}^{+\infty}dq(p^2+q^2)^{\frac{\alpha}{2}}\int_{-\infty}^{+\infty}d\xi\int_{-\infty}^{+\infty}d\eta e^{(ip(x-\xi)+iq(y-\eta))}.$$

The system of Eqs. (1) and (2) gives rise to the *collapse* (catastrophic self-compression of the wave fields), making all solitons unstable, in the case of $\alpha \leq 1$. Families of stable fundamental 2D solitons have been produced in the interval of $1 < \alpha \leq 2$ [12] (the model carries over into the usual 2D $\chi^{(2)}$ system in the limit of $\alpha = 2$). Vortex solitons were found too, but they are completely unstable against spontaneous splitting into a set of moving fundamental solitons. Stable 2D vortex solitons were recently reported as solutions of the 2D fractional nonlinear Schrödinger equation with the cubic-quintic focusing-defocusing nonlinearity [13].

Another new line of the studies of multidimensional optical solitons in various forms makes use of *physics-informed neural networks* (PINN) – in particular, in the application to 2D necklace-shaped soliton chains [14].

Many new theoretical results have been recently reported for 2D and 3D matter-wave solitons in the framework of the Gross-Pitaevskii equations (GPEs) for various species of BEC. In particular, it is well known from theoretical and experimental results that the interplay of long-range dipole-dipole interactions (DDI) with the stabilizing beyond-MF effect gives rise to stable QDs in BEC of magnetic atoms [15,16]. A challenging issue is a possibility of the creation of *stable* QDs with *embedded vorticity* in dipolar BEC. It was known that the dipolar vortex QDs with the simplest isotropic shape, in which the vorticity axis is aligned with the polarization of the atomic dipole moments, fixed by an external magnetic field **B** (Type 1 configuration in Fig. 3), is unstable [17]. Recently, it was demonstrated that an anisotropic vortex-QD configuration, with the vorticity oriented *perpendicular* to the dipoles' polarization (Type 2 state in Fig. 3), may be *stable* [18]. The respective analysis is based on the 3D GPE for the MF wave function $\boldsymbol{\psi}$, written in the scaled form:

$$i\frac{\partial \psi}{\partial t} = -\frac{1}{2}\nabla^2 \psi + g|\psi|^2\psi + \gamma|\psi|^3\psi + \kappa\psi \int U_{\text{DD}}(\mathbf{r}-\mathbf{r}')\,|\psi(\mathbf{r}')|^2 d\mathbf{r}'. \tag{3}$$

Here $g$, $\gamma$ and $\kappa$ are the strengths of the MF local interaction, beyond-MF correction, and DDI, respectively, with the DDI kernel $U_{\text{DD}}(\mathbf{r}-\mathbf{r}') = (1 - 3\cos^2\Theta)/|\mathbf{r}-\mathbf{r}'|^3$, where Θ is the angle between vector $\mathbf{r}-\mathbf{r}'$ and polarizing field **B**, see Fig. 3.

The QD schematically shown as the state of Type 2 in Fig. 3, whose vorticity is perpendicular to field **B**, represents the above-mentioned *stable anisotropic vortex solitons*, such as the prolate one displayed in the top plot of Fig. **4** [18]**.** Equation (3) also produces strongly elongated stable vortex-antivortex-vortex bound states, as shown in the bottom plot of Fig. 4.

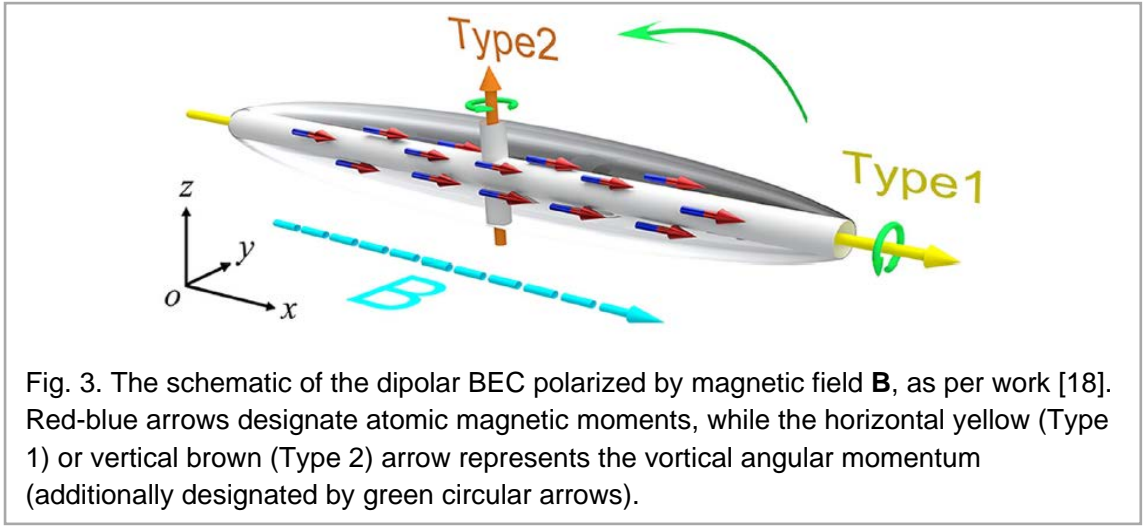

Fig. 3. The schematic of the dipolar BEC polarized by magnetic field **B**, as per work [18]. Red-blue arrows designate atomic magnetic moments, while the horizontal yellow (Type 1) or vertical brown (Type 2) arrow represents the vortical angular momentum (additionally designated by green circular arrows).

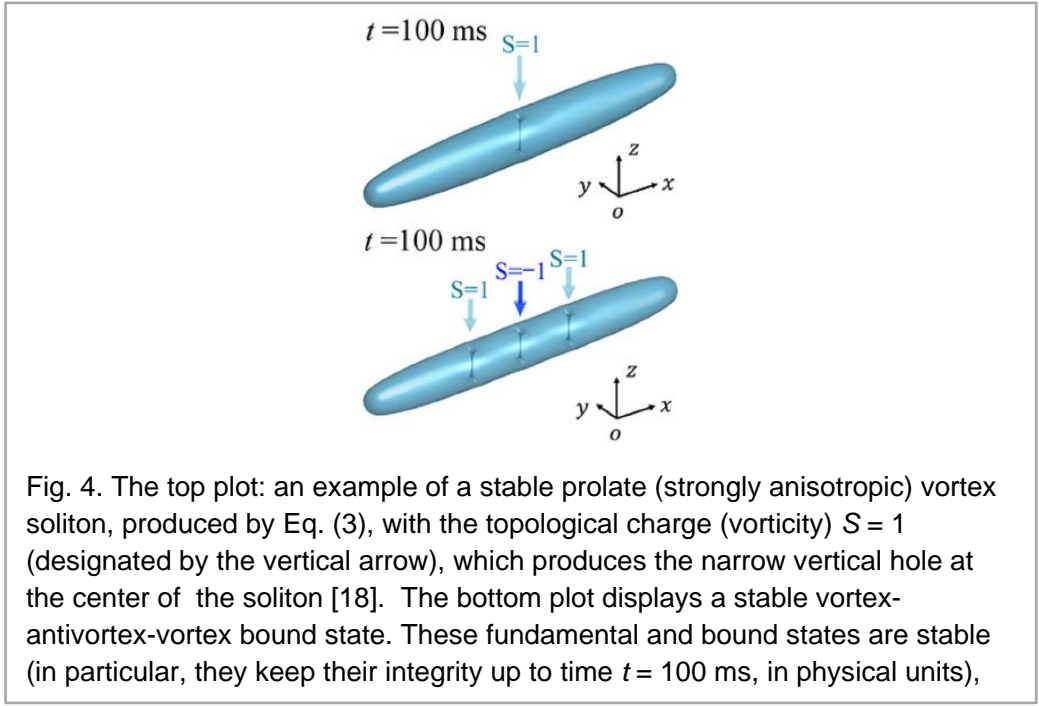

Fig. 4. The top plot: an example of a stable prolate (strongly anisotropic) vortex soliton, produced by Eq. (3), with the topological charge (vorticity) $S = 1$ (designated by the vertical arrow), which produces the narrow vertical hole at the center of the soliton [18]. The bottom plot displays a stable vortex-antivortex-vortex bound state. These fundamental and bound states are stable (in particular, they keep their integrity up to time $t = 100$ ms, in physical units),

Stable 3D solitons, including vortex ones, can also be predicted by GPE with the local self-repulsion and isotropic long-range attraction. The latter term represents artificial gravity, which may be induced by laser illumination of the BEC. The respective 3D GPE is [19]

$$i\frac{\partial \psi}{\partial t} = -\frac{1}{2}\nabla^2\psi + g|\psi|^2\psi - \gamma\psi \int |\psi(\mathbf{r}')|^2|\mathbf{r} - \mathbf{r}'|^{-1}d\mathbf{r}', \qquad (4)$$

with $g > 0$ and $\gamma > 0$. This model readily preduces stable vortex solitons, with the topological charge, at least, up to $S = 6$ [19].

Another physically relevant version of the isotropic nonlocal self-attraction occurs in BEC composed of Rydberg atoms. In that case, the kernel of the respective integral term in GPE is $-C(|\mathbf{r} - \mathbf{r}'|^6 + r_0^6)^{-1}$, with $C > 0$ (instead of $-\gamma|\mathbf{r} - \mathbf{r}'|^{-1}$ in Eq. (4)). The analysis produces stable vortex solitons in a model of this type too [20].

Another noteworthy result was recently reported for the BEC with the local-only interactions, while a variety of nontrivial QD modes was supported by a toroidal trapping potential [21]. The corresponding GPE is

$$i\frac{\partial \psi}{\partial t} = -\frac{1}{2}\nabla^2\psi + g|\psi|^2\psi + \gamma|\psi|^3\psi + U(x,y,z)\psi \qquad (5)$$

(cf. Eq. (3)), where the toroidal potential of radius $r_0$, with radial and vertical widths $d$ and $z_0$, and depth $U_0 > 0$, is defined in the cylindrical coordinates as $U(x,y,z) = -U_0\exp(-(r - r_0)^2/d^2 - z^2/z_0^2)$. Equation (5) gives rise to diverse families of stable soliton-like QDs, including quiescent and rotating necklace-shaped chains of matter-wave solitons and axisymmetric vortices with the topological charge up to 16 (at least).

Furthermore, the application of a torque imparting a perpendicular angular momentum to the vortex QD sets it in a robust precession motion, as illustrated in Fig. 5.

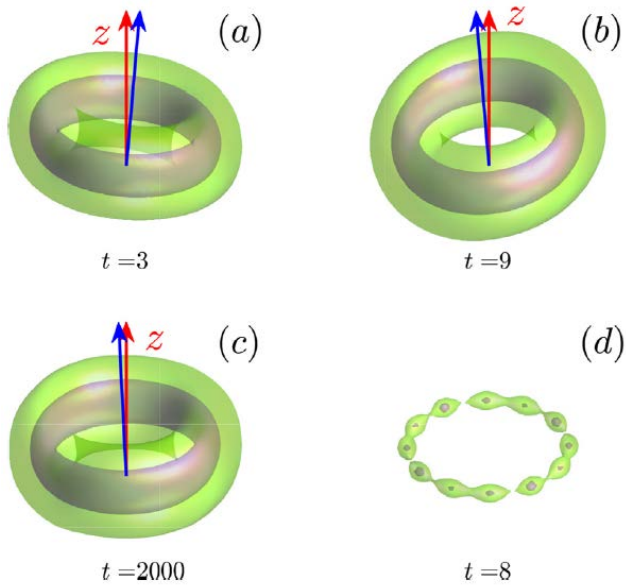

Fig. 5. In panels (*a*)-(*c*) density isosurfaces display, at indicated moments of time, robust precession of the vortex QD, with topological charge 1. This regime of the motion is generated, in the framework of Eq. (5), by the application of an orthogonal torque to a stationary vortex QD [21]. The red and blue arrows indicated, severally, the vertical coordinate axis and instantaneous direction of the vorticity axle. Panel (*d*) shows the result of the development of the collapse in the case when the original vortex QD is unstable.

As concerns other solitons predicted in settings with nontrivial topology, it is relevant to mention recently reported results for 2D matter-wave solitons in BEC with the local self-attraction, placed on the Möbius strip [22].

In conclusion, multidimensional solitons and related nonlinear modes, such as QDs, remain a highly relevant and quickly developing subject for experimental and theoretical studies, in the fields of optics, photonics, quantum matter, and possibly others, such as plasmas, hydrodynamics, and condensed matter, remarkable examples being *skyrmions* in magnetics [23]).

**Acknowledgment**. I thank editors of Advances in Physics for the invitation to submit this Update.